\documentclass[3p]{elsarticle}
\usepackage{hhline}
\usepackage{fancyvrb}
\newcounter{bla}
\usepackage[usenames,dvipsnames]{color}

\journal{Computer Physics Communications}

\begin{document}

\begin{frontmatter}
\title{Algorithm for the replica redistribution in the implementation of parallel annealing method on the hybrid supercomputer architecture}
\author[SCC,HSE]{Alexander Russkov}
\author[HSE]{Roman Chulkevich}
\author[HSE,Landau]{Lev N. Shchur}
\address[SCC]{Science Center in Chernogolovka,142432 Chernogolovka, Russia}
\address[HSE]{National Research University Higher School of Economics, 101000 Moscow, Russia}
\address[Landau]{Landau Institute for Theoretical Physics, 142432 Chernogolovka, Russia}

\begin{abstract}
The parallel annealing method is one of the promising approaches for large scale simulations as potentially scalable on any parallel architecture. We present an implementation of the algorithm on the hybrid program architecture combining CUDA and MPI. The problem is to keep all general-purpose graphics processing unit devices as busy as possible redistributing replicas and to do that efficiently. We provide details of the testing on Intel Skylake/Nvidia V100 based hardware running in parallel more than two million replicas of the Ising model sample. The results are quite optimistic because the acceleration grows toward the perfect line with the growing complexity of the simulated system.
\end{abstract}
\end{frontmatter}

\section{Introduction}
\label{sec-intro}

The development of the large scale simulations is based crucially on the algorithms which are potentially scalable to the billions of processing elements. The big challenge for scientific computing is to develop algorithms and computational frameworks that use the available hardware efficiently. The problem is even more complicated when running code on hybrid supercomputers, combining the conventional CPU with auxiliary accelerating computing devices. 

One of the promising algorithms in the landscape is the parallel annealing algorithm~\cite{Hukushima_Iba,Machta-2010,Machta-2011}, which has two essential and potential features. Firstly, it is fully scalable, and secondly, it can be widely applicable to any system with the partition function. The implementation of the algorithm on the CUDA architecture was recently developed~\cite{Dionicos1,Dionicos2} and applied to the simulation of physical systems with the first-order phase transitions~\cite{Dionicos3}.
The implementation was developed for the single general-purpose graphics processing unit, GPU. The number of GPU cores limits the number of replicas~\footnote{The overall performance of Nvidia GPU is higher when the number of threads is more by order of magnitude than the number of GPU cores.}, which is, therefore, usually of the order of $10^4$. It is known that the systematic errors and statistical errors diminished with an increasing number of replicas, which gives way for unprecedented accuracy of simulations~\cite{Machta-2010,Dionicos2}. The massive parallel realization of the population annealing is the way to simulate a large number of replicas in parallel.  

The population annealing algorithm applies to any system in statistical mechanics formulation, for which the partition function is known, as well as in the optimization problems. Among the published examples are the molecular dynamics~~\cite{PAMD}, glassy fluid~\cite{Amey-Machta}, hard-sphere mixture~\cite{Callaham-Machta}, spin-glasses~\cite{Katz-SG-2018}, and constrained optimization~\cite{Kaboli2017}.

The problem of using several GPU for one task is connected with the problem of efficient redistribution of the replicas between nodes in order to synchronize process of simulation. In the paper we propose and check two algorithms for that. One is the straightforward extension of the original algorithm~\cite{Dionicos1} and the next one is based on the partitioning of replicas, and appears to be more effective.

We organize the paper as follows. We describe the population annealing algorithm briefly in section~\ref{sec-PAmethod}. In section~\ref{sec-MPI}, we introduce algorithms for the redistribution of replicas between different GPU devices. The testing of the combined CUDA/MPI framework presented in the section~\ref{sec-Testing}. Discussion is in the final section~\ref{sec-Discus}.

\section{Population annealing algorithm}
\label{sec-PAmethod}

We discuss here only the essentials of the population annealing (PA) algorithm necessary for understanding the content of the paper. The details of the algorithm can be found in references~\cite{Machta-2010, Dionicos1,Dionicos2}.

The partition function $Z$ of the classical system in the statistical mechanic's approach is

\begin{equation}
Z=\sum_{\left\{i\right\}} \exp\left(-\beta H_i \right),  
\label{part-F}
\end{equation}

\noindent with $H$ is the Hamiltonian, $\beta$ is the inverse temperature, and summation extends over all possible states $i$ of the system.
We refer the reader to the paper~\cite{Kirkpatrick84} for a discussion of the application of simulated annealing in the optimization problem.

The simulation starts with $R_{\beta_0}$ replicas of the system at some value of $\beta_0$. The replica is a particular state of the system, described by the Hamiltonian. The probability $P_j$ of finding the replica at the state $j$ is defined by  

\begin{equation}
P_j=\frac{\exp\left(-\beta_0 H_j \right)}{\sum_{\left\{i\right\}} \exp\left(-\beta H_i \right)}.  
\label{part-F}
\end{equation}

\noindent The simplest and most used way is to choose the initial state at the infinite temperature, $1/\beta_0=0$. In this case, the probabilities $P_j$ are just equal to each other, and one can randomly sample states as equally distributed random variables. We may recommend using one of the libraries~\cite{RNGSSELIB,RNGAVXLIB,PRAND} developed in our group with the universal interface, which allows simple change of the random number generation function~\footnote{Note the difference in realizations of the random number generation (RNG) functions, which are hardware-based, for x86 CPUs with SSE2 extensions~\cite{RNGSSELIB} and AVX extensions~\cite{RNGAVXLIB} of the internal CPU SIMD accelerators, and for external Nvidia GPU accelerators~\cite{PRAND}. The highlight of the libraries is that for all realizations, the same chosen RNG function produces the same sequence of the random numbers given the same initialization of RNG.} and a possibility for the uncorrelated initialization of sample states, using up to $10^{19}$ uncorrelated threads.

The second step is to change the temperature from $1/\beta_0$ to $1/\beta_1$ and calculate the normalized partition function ratio
\begin{equation}
Q(\beta_0,\beta_1)=\frac{\sum_{j=1}^{R_{\beta_0}}\exp\left[-(\beta_1 - \beta_0)E_j\right]}{R_{\beta_0}}
\label{part-Q}
\end{equation}

\noindent and normalized weights 
\begin{equation}
\tau_j(\beta_0,\beta_1)=\frac{\exp\left[-(\beta_1- \beta_0)E_j\right]}{Q(\beta_0,\beta_1)}.
\label{tau}
\end{equation}

The population of replicas at the temperature $1/\beta_1$ generated by resampling the original population at $1/\beta_0$.
We choose the number of replicas in the configuration $j$ from the Poisson distribution keeping the number of replicas close to the original one $R_{\beta_0}$ and taking into account the corresponding weight $\tau_j(\beta_0,\beta_1)$ of configuration $j$ in the resampled population. For the detailed discussion of the possible modifications of this step, we refer readers to the review~\cite{Machta-2011}.

Some replicas with the value of the weights~(\ref{tau}) less than unity will not survive, and others with the most massive weights~(\ref{tau}) will produce more off-springs, as exact copies. Therefore, it is necessary to make them uncorrelated. For that reason, one can use any traditional Markov chain Monte Carlo (MCMC)~\cite{Landau-Binder-book} algorithm for equilibration of the replicas at the new temperature $1/\beta_1$. MCMC will be moving replicas closer to the equilibrium corresponding to the temperature of $1/\beta_1$.  Without the equilibration, the process will correspond to the very fast annealing. Some protocols for choosing the reasonable number of MCMC equilibration time $\theta$ is necessary~\cite{Dionicos1}. We should stress that this step of the algorithm is problem-dependent, and careful choice of the MCMC algorithm should be investigated.

To proceed farther  we change temperature of population to the value $1/\beta_2$ and, calculate weights $\tau_j(\beta_1,\beta_2)$, resample population with $R_{\beta_2}$ replicas, and equilibrate with MCMC. We repeat the annealing process in the same way up to the final temperature of $1/\beta_K$. At each temperature $1/\beta_i$, and after the equilibration, we can compute the thermodynamic observables. The averages computed over the large number of replicas $R_{\beta_i}$.

The nice feature of the population annealing algorithm follows from the form of the partition function ratios~(\ref{part-Q}). The free energy is computed~\cite{Machta-2010, Dionicos3} recursively at all temperatures $1/\beta_i$ by 

\begin{equation}
\beta_i F(\beta_i)=\sum_{l = 1}^{i}Q(\beta_{l-1},\beta_{l})+\ln \Omega,
\label{Free-En}
\end{equation}
\noindent with  $\ln \Omega$ is the value of the free energy at the initial temperature. The practical importance of this recursive relation was used in ref.~\cite{Dionicos3} for the Potts model with first-order phase transition.  Two protocols used, the cooling of the population and the heating of the population. The initial temperature for cooling corresponds to $\beta_0=0$ (at infinite temperature), and the initial temperature of heating is chosen close to the value of $1/\beta_0=0$ (close to zero temperature). The initial value of the corresponding free energy was determined accordingly.

In the rest of the paper, we concentrate on the parallel implementation of the PA algorithm in the hybrid architecture of computing clusters.

\section{Algorithms for redistribution of replicas}
\label{sec-MPI}

For the clarity of discussion, we have to describe our approach using the two-dimensional Ising model, which is the typical approach for discussion of the new algorithms in statistical mechanics. As we mentioned in the Introduction section, the algorithm is suitable for any system in statistical physics and optimization problems with defined goal function.

Hamiltonian of the two-dimensional Ising model on the square lattice with periodic boundary conditions written as

\begin{equation}
H=-\frac{J}{2} \sum\! ^{'} \sigma \sigma'
\label{Ham}
\end{equation}
\noindent with summation extends to all nearest neighbors $\sigma'$ of the spin $\sigma$, coupling constant $J{>}0$, and spins takes values $+1$ and $-1$.

We use the computer code published in~\cite{Dionicos1} with the effective implementation of the PA algorithm for the model~(\ref{Ham}) in CUDA architecture for Nvidia GPU. The authors of the paper~\cite{Dionicos1}  argued that the optimal number of replicas for Nvidia K80 GPU is about 20 000, several times larger than the number of actual cores. We accept the recommendation in our simulations, keeping this number of initial replicas per one GPU card.

Extension of the algorithm to several nodes and several GPU needs additional protocol for the redistribution of the replicas after the replicas resampling step of the PA algorithm, and before the MCMC step. We consider in the paper the MPI architecture for the multi-node multi-GPU realization of hardware.

At the initial step the number of replicas in each node $r^n_{\beta_0}=R_{\beta_0}/N$, $(n=1,2, \ldots, N)$, with $N$ is the total number of GPU cards, and all $r^n_{\beta_0}=20\;480$~\footnote{The optimal number of replicas for Nvidia K80 GPU, according to ref.~\cite{Dionicos1}, is about 20 000. We use the number of replicas multiple of 1024 for the reason of hardware construction.}. Linear size of the square lattice with periodic boundaries is $L=64$. The initial inverse temperature is set to $\beta_0=0$, the final value $\beta_K=1$, with $K=200$, so the annealing step is $\Delta\beta=0.005$. 

In the most straightforward possible realization, all $N$ nodes run independently, keeping the global normalization of the partition function ratio according to the expression~(\ref{part-Q}), without any redistribution of the replicas between nodes. It is not the efficient realization leading to the case with one node of the nodes will run most of the replicas due to the large fluctuations in the number of the off-springs. We do not report any details of this possible realization in the paper.

To approximately equalize the workload of GPU cards, we consider two algorithms for the redistribution of replicas between nodes. MPI library functions allows the direct exchange of GPU memories.

First, we consider the algorithm (we name it the ``naive'' algorithm), considering the current number of replicas and the number of offsprings at each node. The number of replicas for redistribution is proportional to the number of offspring at each node. The central node collects the number of replicas at each node, calculates the total current number of replicas $R_{\beta_i}$, and sends this number to the nodes as the normalization factor in the partition function ratios.

\begin{figure}[ht] \center
\includegraphics[width=.5\columnwidth]{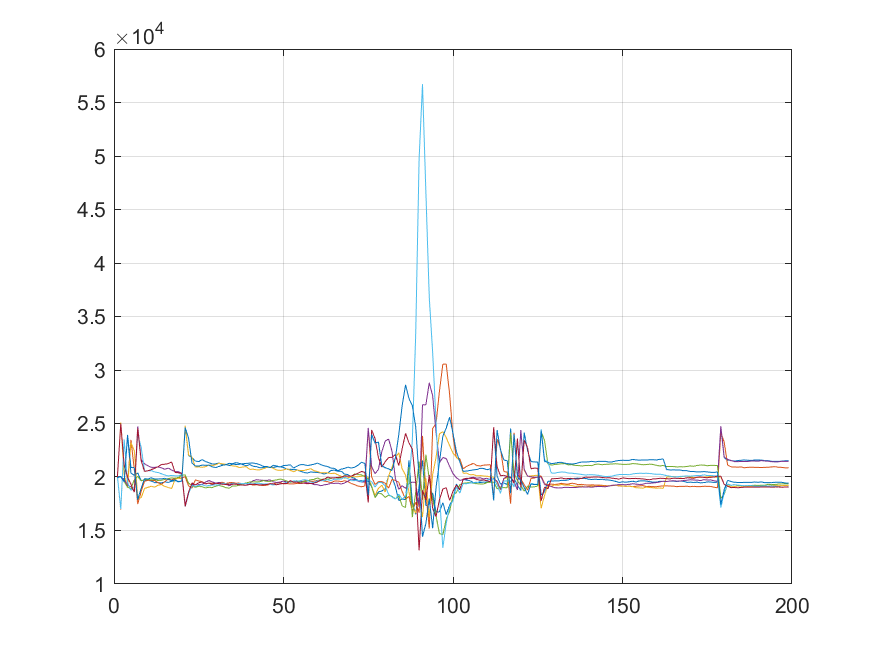}\includegraphics[width=.5\columnwidth]{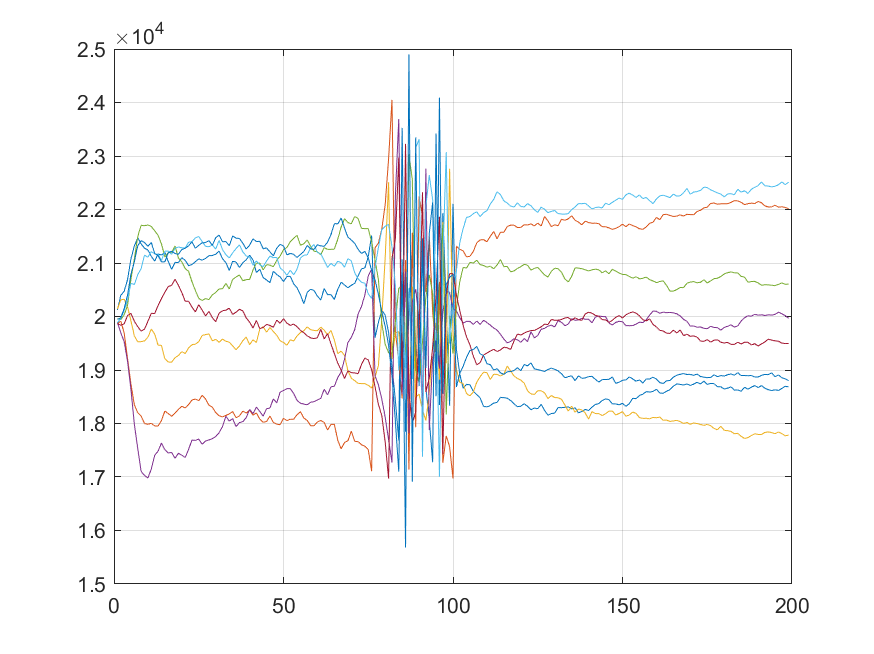}
\caption{The number of replicas at each of eight GPU cards (marked with the different colors) at 200 annealing steps. Left: the first (``naive'') algorithm. Right: the main algorithm.}
\label{fig:red-1}
\end{figure}

The ``naive'' algorithm shows satisfactory behavior at each temperature step and not in the critical region, as one can see from the left panel of Fig.~\ref{fig:red-1}, which shows the ``temporal'' change in the number of replicas at all eight GPU cards ($N=8$). The fluctuations in the number of replicas  $r^n_{\beta_i}$ for the temperature steps $i\approx 5$ and 20,  relaxes fast to the flat behavior. The system behaves similarly also in the areas of temperature steps $i$ around 110, 130, and 180. The drastically different behavior happens in the vicinity of the phase transition, which corresponds to a temperature step $i\approx 90$. In this region, fluctuation in one of the replicas reaches a value larger than 55 thousand (almost three times larger than $r^n_{\beta_0}$).

The main algorithm is more sophisticated, and it uses partitioning of the replicas in the blocks. At a given node and at each annealing step we have the current number of replicas $r^n_{\beta_0}$ and the ``desired'' number of replicas $r_n$. 

Replicas redistributed by blocks. Each block contains 1024 replicas~\footnote{The number of threads per block for the parallel algorithm Nthreads=1024 or 2048 for CUDA capability 2.0 and above and Nthreads=512 for CUDA capability 1.x).} , and in the algorithm realization, the total number of blocks is ${\sf nblc}=R_{\beta_0}/1024$. We define the window for the aloud number of blocks at each GPU card, with the desired number of replicas at each card is $r_n$. We allow the difference in the number of blocks at each node fixed by the parameter $\sf maxexc$, we use in simulations ${\sf maxexc}=1$, and we found not much difference while using ${\sf maxexc}=2$. Before the redistribution step,  algorithm calculates the excess value at each node ${\sf maxg}_n=r^n_{\beta_i}-r_n$ or the shortage value ${\sf minl}_n=r_n-r^n_{\beta_i}$, depending on case which one is positive.

The number of replicas during the simulation is flatter for the main algorithm, as can be seen from the right panel of Fig.~\ref{fig:red-1}. Not the higher level of fluctuations in the critical region, which is instead of moderate amplitude. We present simulations with the constant annealing step $\Delta\beta$. Fluctuations can be made smaller and less frequent using the adaptive temperature step, as discussed in the papers~\cite{Dionicos1}, thus making simulations more effective.

\begin{figure}[ht] \center
\includegraphics[width=.5\columnwidth]{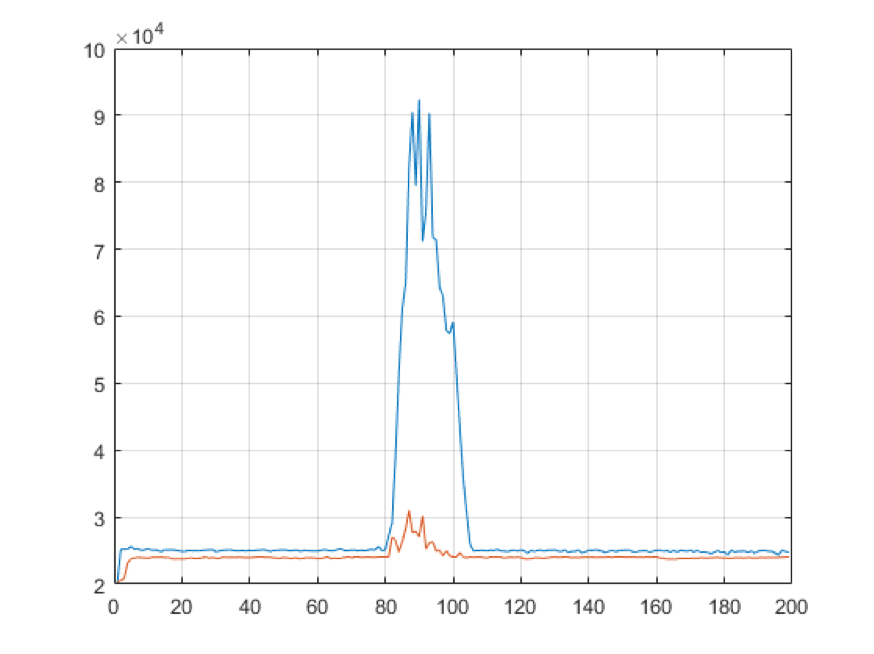}
\caption{The  maximum number of replicas at the node at 200 annealing steps.}
\label{fig:comp12}
\end{figure}

Figure~\ref{fig:comp12} shows the {\it maximum number} of the threads at one GPU card (the card with the maximum numbers of threads chosen at each temperature step) at different annealing steps comparing the ``naive'' and main algorithm for the typical simulation. In the critical region, the number of replicas at one GPU reaches as much as 90 thousand with the ``naive'' algorithm, 4.5 times larger than the desired number $r_n$. Other GPUs have to wait for the completion of the simulations at this overloaded node. Accordingly, the total time of simulation represented in Fig.~\ref{fig:comp12} is longer by factor 1.45 for the ``naive'' algorithm in comparison with the main algorithm. The maximum number of replicas for the main algorithm is at a reasonable level of magnitude.

In the following, we use the main algorithm for the analysis of the parallel population algorithm scalability.

\section{Testing of scalability}
\label{sec-Testing}

\subsection{Simulation task}

We simulate a square lattice Ising model with a lattice size $L=64$ and use protocol with the constant inverse temperature step $\Delta\beta=0.005$, the initial inverse temperature $\beta_0=0$, and the final inverse temperature $\beta_{200}=1$. The number of steps for Monte-Carlo (MCS) equilibration is $\theta=1$, so the most time of simulations is spending for the population re-weighting and redistribution of replicas. We have to note that such a small value of MCMC is the most extreme case for the MPI realization. The more time simulation will be spent on the MCMC, the less relative time would be spent on the redistribution of replicas, which is one of the bottlenecks in the MPI realization of simulations..

We should stress that the small number of MCMC steps $\theta=1$ we use in the testing is not sufficient for any realistic simulations. The reason for the smallest number of MCMC steps is that simulations of the simple Ising model with the minimum number of MCMC steps will give us a minimal estimation of scalability. In the last subsection, we demonstrate how much scalability becomes better for the same task and with the realistic number of MCMC steps $\theta=10$ for the Ising model simulations.

We test our algorithm in the two computing environments, the Manticore cluster, and the SCC cluster. The Manticore cluster is based on the nodes with two Intel Xeon E5-2683 v3 CPU at 2.1 GHz and with 4 Nvidia Tesla K80 GPU each. The SCC cluster is based on the nodes with two Intel Xeon Gold 6152 2.1GHz CPU, with onboard memory DDR4 2.666GHz 768GB RAM, and with 4 GPU Tesla V100 32Gb. It is the two most typical computing configurations nowadays.

\subsection{Small scale testing at Manticore}

In order to calibrate performance with other published works, we use the same computer facility as in paper~\cite{Dionicos1}, the two nodes of the Manticore cluster with 8 GPU cards.  We use single spin coding and the Ising model with a square lattice with a size $L=64$. Simulations demonstrate speedup in 246 times of the single K80 in comparison with single CPU~\cite{Dionicos1}. The initial number of replicas per GPU is fixed, and it is 20 blocks with 1024 replicas each, 20480 replicas per GPU.

Table~\ref{tab:mc-nodes} present the time of simulation of the task, varying the number of GPUs. The left panel of the table shows the computing time while using a different number of GPUs at a single node. The total time grows by 39 percent with the increasing number of GPUs from 1 to 4, and 4 times more replicas were running in parallel. So, the total simulation is faster in the wall clock, with factor 2.87. 

The right panel of the table~\ref{tab:mc-nodes} shows the computing time while using an equal number of GPUs at both nodes. In this case, the total time grows by 33 percent with the increasing number of GPUs from 2 to 8, and 4 times more replicas were running in parallel. So, the total simulation is faster in the wall clock, with factor 3. 

Comparing the left and right panels in Table~\ref{tab:mc-nodes} one can see that running GPUs at different nodes is faster by 12 percent than running the same number of GPUs at the single node.

\begin{table}[ht]
\center
\begin{tabular}{|c|c|}
\hline
Number of GPU &  time in sec \\
\hline
1 & 707.6 \\
2 & 856.2 \\
4 &  983.7\\
& \\
\hline
\end{tabular}
\begin{tabular}{|c|c|}
\hline
Number of GPU &  time in sec \\
\hline
& \\
2=1+1 & 765.3 \\
4=2+2 & 884.6 \\
8=4+4 &  1021.5 \\
\hline
\end{tabular}
\caption{The left table: The simulation time on the single node with 4 GPU. The right table: The time simulation on the two nodes with 4 GPU each, and symmetric number of GPU per node. In all cases, the initial number of replicas per GPU is 20480 -- simulations on Intel Xeon E5-2683 v3 CPU and Nvidia Tesla K80.}
\label{tab:mc-nodes}
\end{table}

We simulate about 163480 replicas in parallel on the Manticore hardware in this subsection. Combining with the result of paper~\cite{Dionicos1} on the single CPU, we conclude the total simulation is more productive by factor 1476 in comparison with the single CPU.

\subsection{Small scale testing at SCC}

We perform the same simulations as in the previous subsection on the SCC cluster with 26 nodes, with 104 GPUs available. In simulations, we use OPENMP 4.0.1, CUDA version 10.2, and NVIDIA driver version 440.33.01.

\begin{table}[ht]
\center
\begin{tabular}{|c|c|}
\hline
Number of GPU &  $T_{s,1}, sec $ \\
\hline
1 & 153.7 \\
2 & 222.5 \\
4 &  343.2\\
& \\
\hline
\end{tabular}
\begin{tabular}{|c|c|}
\hline
Number of GPU &  $T_{d,1}, sec$ \\
\hline
& \\
2=1+1 & 214.2 \\
4=2+2 & 305.9 \\
8=4+4 &  442.5 \\
\hline
\end{tabular}
\caption{The left table: The simulation time $T_{s,1}$ on the single node with 4 GPU. The right table: The time simulation $T_{d,1}$ on the two nodes with 4 GPU each, and symmetric number of GPU per node. In all cases, the initial number of replicas per GPU is 20480 -- simulations on Intel Xeon Gold 6152  CPU and Nvidia Tesla V100. Number of MCS for equilibration $\theta=1$.}
\label{tab:skk-nodes-1}
\end{table}

Table~\ref{tab:skk-nodes-1} presents the time of simulation of the task, varying the number of GPUs. The left panel of the table shows the computing time while using a different number of GPUs at a single node. The total time grows by factor 2.23 with the increasing number of GPUs from 1 to 4, and 4 times more replicas were running in parallel. So, the total simulation is more productive, with factor 1.79. 

The right panel of the table~\ref{tab:skk-nodes-1} shows the computing time while using an equal number of GPUs at both nodes. 
The total time grows by factor 2.07 with the increasing number of GPUs from 2 to 8, and the total simulation is more productive, with factor 1.94.

Comparing results obtained with the Manticore cluster and SCC cluster, we found that the productivity grows slower on the SCC cluster, 55-60 percent worse than on the Manticore cluster. Furthermore, the total time is better on the SCC cluster, with a coefficient from 1.4 to 2.3. 

The speed of simulation on V100 is higher than on K80, and the memory channels much frequently loaded with memory IO operations. In the case, the GPU cards will spend more time on simulations, and less time on the redistribution of replicas, the total productivity should grow better for V100 cards.

For the next step of simulations, we use a more realistic number of MCMC steps $\theta=10$ for the Ising model simulations. This way, we decrease the relative time of simulations spent on the replica redistribution step. 

Table~\ref{tab:skk-nodes-10} presents the time of simulation of the task with $\theta=10$ MCS. The left panel of the table shows the computing time while using a different number of GPUs at a single node. The total time grows by factor 1.26 with the increasing number of GPUs from 1 to 4, and 4 times more replicas were running in parallel. So, the total simulation is more productive, with factor 3.18. The right panel of the table~\ref{tab:skk-nodes-10} shows the computing time while using an equal number of GPUs at both nodes. The total time grows by factor 1.25 with the increasing number of GPUs from 2 to 8, and the total simulation is more productive, with factor 3.21.

The third columns in the Table~\ref{tab:skk-nodes-10} demonstrate how the time of simulations changed with increasing number of GPUs, dividing the time $T_{s,10}$ and $T_{d,10}$ in the second columns of the Table~\ref{tab:skk-nodes-10} onto the corresponding times $T_{s,1}$ and $T_{d,1}$ from the second columns of the Table~\ref{tab:skk-nodes-1}. The good sign is that the efficiency of the realistic simulations with $\theta=10$ growing with the number of GPU used.

\begin{table}[ht]
\center
\begin{tabular}{|c|c|c|}
\hline
Number of GPU &  $T_{s,10}$, sec & $T_{s,10}/T_{s,1}$\\
\hline
1 & 349.0 & 2.27 \\
2 & 383.0  & 1.72 \\
4 &  438.9 & 1.27\\
& \\
\hline
\end{tabular}
\begin{tabular}{|c|c|c|}
\hline
Number of GPU &  $T_{d,10}$, sec & $T_{d,10}/T_{d,1}$ \\
\hline
& \\
2=1+1 & 368.8 & 1.72 \\
4=2+2 & 398.6 & 1.30 \\
8=4+4 &  459.3 & 1.04 \\
\hline
\end{tabular}
\caption{The left table: The simulation time $T_{s,10}$ on the single node with 4 GPU. The right table: The time simulation $T_{d,10}$ on the two nodes with 4 GPU each, and symmetric number of GPU per node. In all cases, the initial number of replicas per GPU is 20480 -- simulations on Intel Xeon Gold 6152  CPU and Nvidia Tesla V100. Number of MCS for equilibration $\theta=10$.}
\label{tab:skk-nodes-10}
\end{table}

\subsection{Large scale testing}

We simulate our task with the number $\theta=10$ of MCS on the SCC cluster.

\begin{figure}[ht] \center
\includegraphics[width=.5\columnwidth]{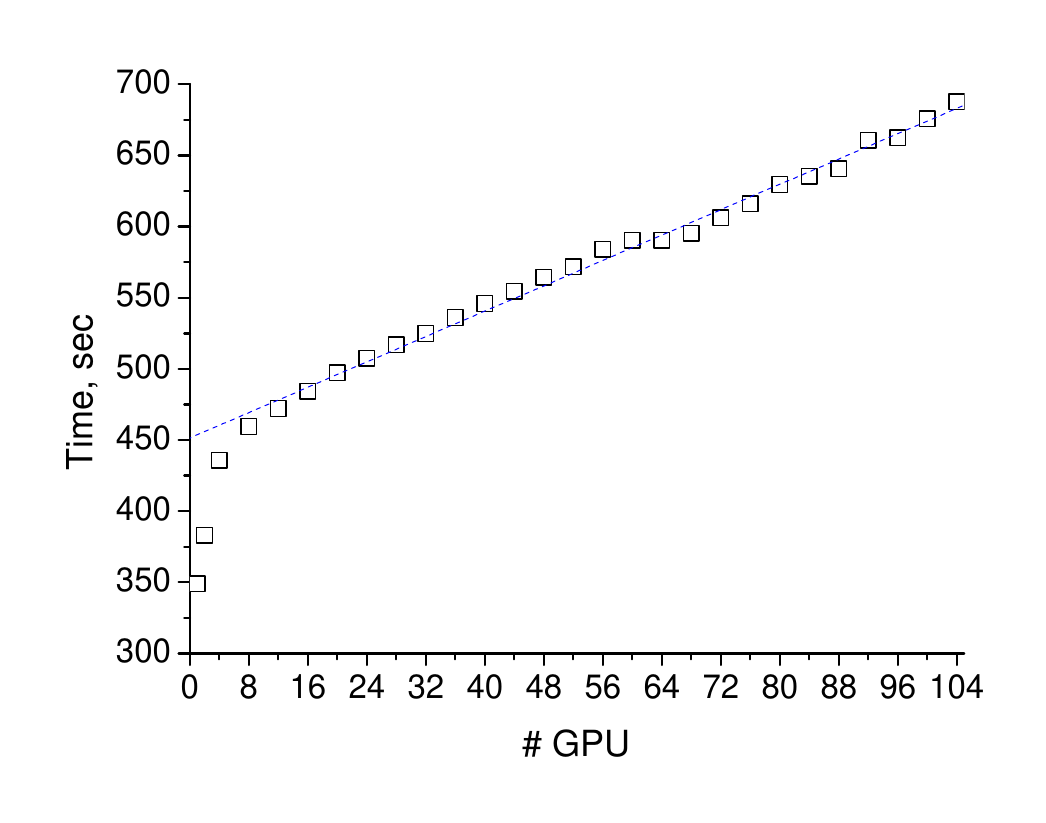}~\includegraphics[width=.5\columnwidth]{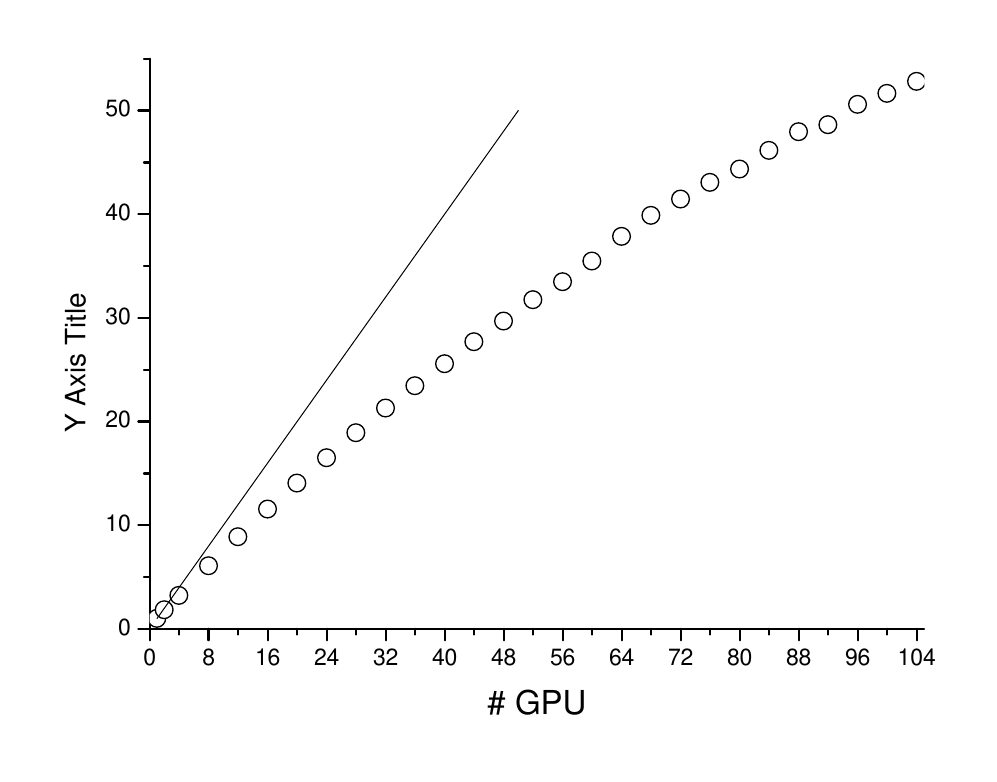}
\caption{The left panel: The time in seconds of the task as a function of the number of GPUs. The dotted line has slope 2.2. The right panel: Acceleration of the replica simulation with the number of GPUs. The number of MCS $\theta=10$ for equilibration. The line shows acceleration without replica redistribution between GPU.}
\label{fig:time26-10}
\end{figure}

In the left panel of Figure~\ref{fig:time26-10}, we show time dependence of the whole task with the number of GPU cards from 1 to 104. The simulation time grows because of the redistribution of the replicas between GPUs. At the same time, the total number of replicas is growing 104 times. We can plot data taking into account the growth of the task, and calculate the acceleration as the time of simulation of the task on $N$ GPU cards divided by the time of simulation on one GPU card -- the resulting data presented in the right panel of Figure~\ref{fig:time26-10}. More precisely, the acceleration is the speed of the replica simulation divided by the unit of time. The acceleration grows with about 52 times as we use all 104 GPUs of the SCC cluster, i.e., 50 percent of the ideal acceleration.

One can ask why we do not use 104 GPUs separately and get the 100 percent of acceleration? It would be the case of 104 independent simulations of the population annealing algorithm, i.e. simulations with 104 independent populations, and the weighted averages will be calculated within the each population separately. 

The answer is that with the weighted averaging, the population annealing algorithm gives a very precise estimation of the measured quantities for a large population size $R$. In our case, simulating 2 129 920 replicas of a single population, we reduce systematic error~\cite{Machta-2010} by a factor $1/R_{\beta_0}$, which is smaller than $10^{-6}$.

The equilibration process also depends not only on the number of Monte Carlo steps but on the complexity of calculation at each step as well. Simulating more complex problem which involves more operations with the calculation of the weights and more operations to calculate averages of interest than with the Ising model, will have less relative time spent on the redistribution of replicas; and lead to a better acceleration than in the presented example.

\section{Discussion}
\label{sec-Discus}

We propose a parallel version of the population annealing algorithm using the block redistribution scheme of replicas. We test the algorithm with the square lattice Ising model.  The results are quite optimistic -- the more complicated model, the better acceleration of the problem. 

The motivation for using the large population size is that with the larger population $R$, one can achieve better accuracy of simulations. The statistical error is reduced with factor $1/\sqrt{R}$ and systematic error reduce even faster as $1/R$ while using the reweighing procedure~\cite{Machta-2010, Dionicos1}.  The presented massive parallel population annealing MPI/Cuda approach can be used for the complex systems. As an example, it gives a natural way to calculate the distribution of the order parameter for the spin-glass model~\cite{SG} at each annealing step, averaging over the population of a large population $R$ of replicas. The approach can be helpful for complex optimization problems as well.

In summary, we present the version of the population annealing algorithm for the load balancing of GPU cards, keeping an approximately equal number of replicas per the card, thus optimizing the load balance. The algorithm minimizes the extensive memory exchange between nodes using blocks of replicas. We simulated up to 2 million replicas in one run on the 104 GPU computation cluster.

\section{Acknowledgments}

This work has been initiated under the grant 14-21-00158  and finished within the framework of the grant 19-11-00286, both from the Russian Science Foundation.
We use for the small scale testing the Manticore cluster of ANR laboratory at Science Center in Chernogolovka and for the large scale testing the supercomputing facility of the National Research University Higher School of Economics. Special thanks to Pavel Kostenetskiy for the support of the full-scale simulations.


\frenchspacing

\begin{thebibliography}{99}

\bibitem{Hukushima_Iba} K. Hukushima, Y. Iba, AIP Conf. Proc. {\bf 690} (2003) 200
\bibitem{Machta-2010} J. Machta, Phys. Rev. E {\bf 82} (2010) 026704
\bibitem{Machta-2011} J. Machta and E.S. Ellis, J. Stat. Phys. {\bf 144} (2011) 541
\bibitem{Dionicos1} L.Yu. Barash, M. Weigel, M. Borovsk\'y, W. Janke, and L.N. Shchur, Comp. Phys. Commun. {\bf 220} (2017) 341
\bibitem{Dionicos2} L. Shchur, L. Barash, M. Weigel, and W. Janke, Comm. in Comp. and Inform. Science {\bf 965}  (2019) 354
\bibitem{Dionicos3} L.Yu. Barash, M. Weigel, L.N. Shchur, and W. Janke, Eur. Phys. J. Spec. Top. {\bf 226} (2017)  595
\bibitem{PAMD} H. Christiansen, M. Weigel, and W. Janke, Phys. Rev. Lett. {\bf 122} (2019) 060602
\bibitem{Amey-Machta} C. Amey and J. Machta, Bull. Amer. Phys. Soc. {\bf 65} (2020) D45.00009
\bibitem{Callaham-Machta} J. Callaham and J. Machta, Phys. Rev. E {\bf 95} (2017) 063315
\bibitem{Katz-SG-2018} A. Barzegar, C. Pattison, W. Wang, and H.G. Katzgraber, Phys. Rev. E {\bf 98} (2018) 053308
\bibitem{Kaboli2017} S.Hr. Aghay Kaboli, J. Selvaraj, and N.A. Rahim, J. Comp. Sci. {\bf 19} (2017) 31
\bibitem{Kirkpatrick84} S. Kirkpatrick, J. Stat. Phys. {\bf 34} (1984) 975
\bibitem{RNGSSELIB} L.Yu. Barash and L.N. Shchur, Comput. Phys. Commun., {\bf 182} (2011) 1518; L.Yu. Barash, L.N. Shchur, Comput. Phys. Commun., {\bf 184} (2013) 2367
\bibitem{RNGAVXLIB} M.S. Guskova, L.Yu. Barash, and L.N. Shchur, Comp. Phys. Commun., {\bf 200} (2016) 402
\bibitem{PRAND} L.Yu. Barash and L.N. Shchur,  Comput. Phys. Commun., {\bf 185} (2014) 1343 
\bibitem{Landau-Binder-book} D.P. Landau and R. Binder, {\it A Guide to Monte Carlo Simulations in Statistical Physics} (Cambridge University Press, Cambridge, 2014)
\bibitem{SG}  M. M\'ezard, G. Parisi, and M.A. Virasoro, {\it Spin Glass Theory and Beyond} (World Scientific, Singapore, 1987)
\end{thebibliography}
\end{document}